\documentclass[10pt, pra, aps,twocolumn, showpacs,floatfix, longbibliography, superscriptaddress]{revtex4-2}

\setcitestyle{super}
\usepackage{amsmath,amssymb,physics}
\usepackage[utf8]{inputenc}
\usepackage{datetime}
\usepackage{bm}
\usepackage{enumitem}
\usepackage{graphicx}
\usepackage{braket}
\usepackage{xcolor}
\usepackage{MnSymbol}
\usepackage[T1]{fontenc} 
\usepackage[normalem]{ulem}
\usepackage{mathtools}
\usepackage{bbold}
\usepackage{siunitx}
\usepackage[hypertexnames=false]{hyperref}
\usepackage{soul}
\usepackage{gensymb}
\usepackage[dvipsnames]{xcolor}
\usepackage{xfrac}
\usepackage{tikz}

\renewcommand{\selectlanguage}[1]{}

\hypersetup{
    colorlinks,
    linkcolor={red!50!black},
    citecolor={blue!50!black},
    urlcolor={blue!80!black}
}

\begin{document}

\makeatletter
\def\@fnsymbol#1{\ensuremath{\ifcase#1\or \dagger\or  *\or
   \mathsection\or \mathparagraph\or \|\or **\or \dagger\dagger
   \or \ddagger\ddagger \else\@ctrerr\fi}}
\makeatother

\title{Quantum Resistance Paradox of Low-Dimensional Superfluids}

\author{Simon Wili}
\affiliation{Institute for Quantum Electronics and Quantum Center, ETH Z\"urich, 8093 Z\"urich, Switzerland}
\author{Meng-Zi Huang}
\altaffiliation{Co-Corresponding Author: mzhuang@mail.ecnu.edu.cn; New affiliation: Institute of Quantum Science and Precision Measurement, East China Normal University, Shanghai 200062, China}
\affiliation{Institute for Quantum Electronics and Quantum Center, ETH Z\"urich, 8093 Z\"urich, Switzerland}
\author{Tommaso Bonaccorsi}
\affiliation{Institute for Quantum Electronics and Quantum Center, ETH Z\"urich, 8093 Z\"urich, Switzerland}
\author{Michael Mühlematter}
\affiliation{Institute for Quantum Electronics and Quantum Center, ETH Z\"urich, 8093 Z\"urich, Switzerland}
\author{Mohsen Talebi}
\affiliation{Institute for Quantum Electronics and Quantum Center, ETH Z\"urich, 8093 Z\"urich, Switzerland}
\author{Yaakov Yudkin}
\affiliation{Institute for Quantum Electronics and Quantum Center, ETH Z\"urich, 8093 Z\"urich, Switzerland}

\author{Alex Gómez-Salvador}
\affiliation{Institute for Theoretical Physics, ETH Z\"urich, 8093 Z\"urich, Switzerland}
\author{Filip Marijanovic}
\affiliation{Institute for Theoretical Physics, ETH Z\"urich, 8093 Z\"urich, Switzerland}
\author{Eugene Demler}
\affiliation{Institute for Theoretical Physics, ETH Z\"urich, 8093 Z\"urich, Switzerland}

\author{Tilman Esslinger}
\altaffiliation{Co-Corresponding Author: esslinger@ethz.ch }
\affiliation{Institute for Quantum Electronics and Quantum Center, ETH Z\"urich, 8093 Z\"urich, Switzerland}

\date{\today}

\begin{abstract}

Resistance in standard conductors decreases with increasing cross-section. Yet, in low-dimensional superconductors and superfluids residual resistance arises from topological fluctuations of the order parameter manifesting as phase slips in one-dimensional (1D) and vortices in two-dimensional (2D) systems. How resistance and dissipation evolve as geometry interpolates between these regimes remains an open question. This evolution is masked in solid-state experiments by disorder, impurities, and geometric imperfections,\cite{nano8060442,PhysRevB.77.104516,ARUTYUNOV20081} and poses theoretical challenges due to competing dissipative processes and pronounced finite-size effects. Here, we use a defect-free unitary Fermi gas in a digitally programmable transport geometry to isolate geometric effects on superfluid dissipation and discover a paradox: in the crossover from 1D to 2D, the resistance reaches a minimum. There, widening a channel increases its resistance. Narrower, quasi-1D channels show dissipation described by Langer-Ambegaokar-McCumber-Halperin theory of phase slips.\cite{doi:10.1142/S021797921005644X} In this regime, varying the channel width yields the predicted exponential scaling of the activation factor over more than ten orders of magnitude. Wider, quasi-2D channels show dissipation consistent with a finite-size vortex model.\cite{nakagawa_vortex_2024} The minimal dissipation in the dimensional crossover reflects a transition in the dominant dissipative mechanism, with both phase slips and vortices simultaneously suppressed. Our measurements suggest a route to minimizing dissipation in superconducting devices and provide a benchmark for theoretical efforts aimed at describing the dimensional crossover.
\end{abstract}
\maketitle

\begin{figure*}[t]

    \centering
\includegraphics[width=\textwidth]{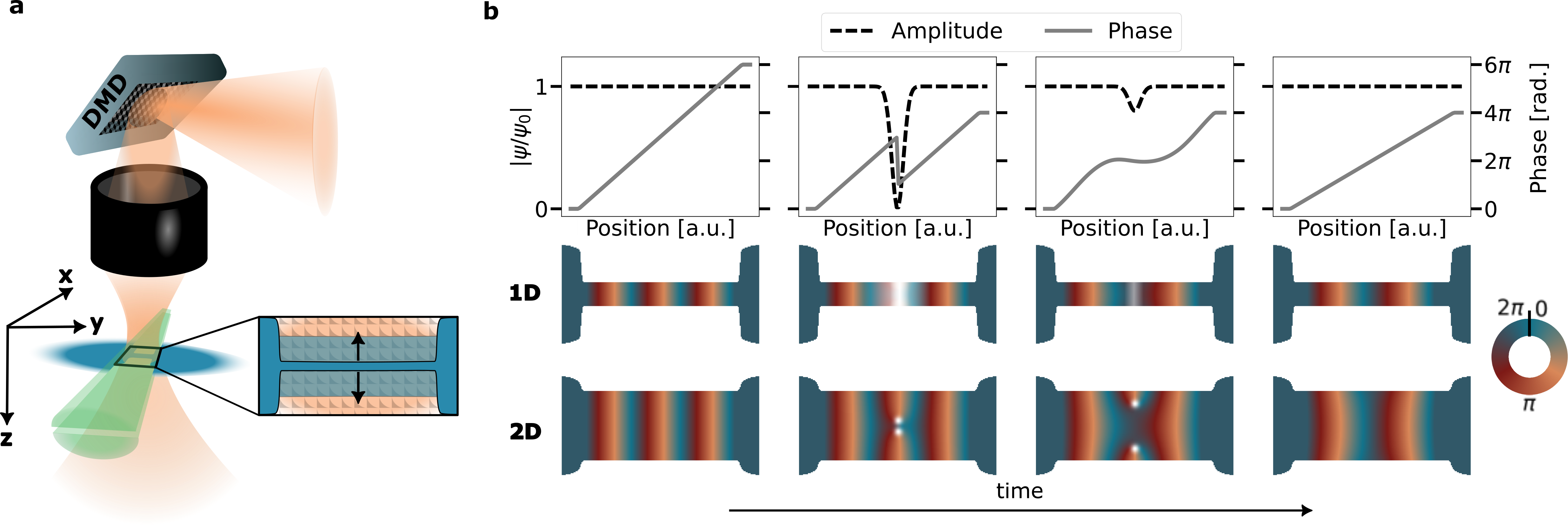}
    \caption{\textbf{Illustration of the setup and the dissipative processes in one and two dimensions}\newline \textbf{a}: A transport channel of programmable width connects two superfluid reservoirs (blue). The channel is formed by two repulsive beams (green and orange), one of which is tunable by a DMD. \textbf{b}: Dissipation in a narrow, quasi 1D channel occurs due to phase slips, where a depletion of the order paramete allows for a relaxation of the phase gradient and kinetic energy (top row). The corresponding phase (color coded) evolution of a phase slip in a narrow channel is plotted in the central row. In a wide, quasi 2D channel, dissipation is governed by vortex-antivortex pairs traversing the channel (bottom row).
    }
    \label{fig: fig1}       
\end{figure*}
The Landau paradigm provides a powerful framework for understanding many phases of matter through the lens of spontaneous symmetry breaking.\cite{Ginzburg:1950sr, 10.21468/SciPostPhysLectNotes.11} In superfluids and superconductors, this breaking manifests in the emergence of a macroscopic condensate which exhibits long-range phase coherence. This 
coherence is ultimately responsible for properties such as dissipationless transport,\cite{10.1098/rspa.1935.0048,PhysRev.60.356} the Josephson effect,\cite{JOSEPHSON1962251} and in the case of superconductivity the Meissner effect.\cite{Meissner1933} While relatively robust in bulk systems, in lower-dimensional systems, the enhanced effect of fluctuations -- and in particular topological defects -- can suppress this long-range phase coherence and govern superfluid transport.\cite{Tinkham1996,doi:10.1142/S021797921005644X}

For example in a voltage-biased superconducting wire the applied voltage continuously winds the superconducting phase, causing an increase of the phase gradient between the contacts. The supercurrent, proportional to this gradient, would grow without bound unless intermittent phase fluctuations relax the accumulated gradient. 
By relaxing the gradient, the fluctuations dissipate energy, thereby causing a resistance.\cite{GIODANO1994460, ARUTYUNOV20081} The characteristics of the fluctuations and their corresponding resistance depend on the dimensionality of the setup as follows.

In one-dimensional (1D) wires, discrete relaxation events known as phase slips dominate the dymanics. Microscopically, a phase slip occurs when the magnitude of the superconducting order parameter is transiently and locally suppressed over a healing length $\xi$. Where the amplitude vanishes, the phase is ill-defined, allowing for phase jumps of $2\pi$. As the order parameter recovers, the wire returns to a state with a uniform amplitude but a reduced phase gradient (Fig.~1b, top row).\cite{doi:10.1142/S021797921005644X} 

In two-dimensional (2D) geometries, such as thin films, dissipation instead arises from the nucleation and motion of topological defects known as vortices. A vortex is a point-like singularity of the phase around which the phase winds by $2\pi$, with a small core where the amplitude is suppressed. Driven by a current, vortices move across the film and dissipate energy. Each time a vortex traverses the sample, the phase difference between the contacts relaxes by $2\pi$. Repeated vortex crossings generate a finite voltage and hence an effective resistance (Fig.~1b, bottom row).\cite{Halperin1979, RevModPhys.59.1001}
In intermediate, thin strip geometries, both vortices and phase slips contribute to dissipation.

Previous studies included experiments on superconducting rings,\cite{petkovic_deterministic_2016} Dayem-bridge nanowires,\cite{skocpol_phase-slip_1974,PhysRevB.70.214506,PhysRevApplied.19.044073,PhysRevB.106.134518} Bose–Einstein condensates in dumbbell, \cite{eckel_contact_2016,gauthier_quantitative_2019} cylindrical,\cite{navon_emergence_2016} or toroidal geometries,\cite{WeakLinkPS} optical lattices,\cite{LatticePhaseSlip,PhysRevLett.86.4447} and fermionic Josephson junctions.\cite{RoatiJosephsonBECBCS,RoatiJosephsonPhaseSlip,MoritzIdealJosephson} They have explored various facets of phase slip and vortex physics, including direct imaging of vortices,\cite{zwierlein_vortices_2005,RoatiJosephsonPhaseSlip,PhysRevLett.83.2498,PhysRevLett.84.806} the BKT phase transition,\cite{BKTExperiment} and the recent observation of Shapiro steps in atomic Josephson junctions.\cite{doi:10.1126/science.ads9061,doi:10.1126/science.ads8885} However, the experimental challenge of isolating and characterizing geometric or dimensional effects on dissipation remains unresolved. 

With the rise of quantum simulation of devices based on highly controllable ultracold atomic gases,\cite{Krinner_2017,RevModPhys.94.041001} it is now possible to shed new light on how these dissipative processes emerge in quantum systems.

In this work, we employ ultracold, strongly interacting fermions in a two-terminal configuration to study nonlinear transport through a digitally programmable channel. The optically shaped channel can be reproducibly tuned from a quasi-1D to a quasi-2D geometry, providing a platform ideally suited for probing the influence of geometry on fluctuation-driven dissipation in the absence of defects or impurities.

In the narrow, one-dimensional channel limit we identify signatures of phase slips, in agreement with the Langer-Ambegaokar-McCumber-Halpering (LAMH) theory.\cite{LA_original,MH_original,doi:10.1142/S021797921005644X} In particular, we observe the predicted exponential suppression of the LAMH activation factor over roughly ten orders of magnitude upon widening the channel. For even larger widths, the system crosses over to a two-dimensional regime in which dissipation is dominated by vortices, as indicated by the observed power-law dependence of the current–voltage characteristics predicted for vortex dynamics. 

Beyond establishing these limiting behaviors, our measurements provide a direct window into the crossover between one- and two-dimensional transport, a regime where analytical modeling is challenging. It is in this crossover, where widening the channel paradoxically increases its resistance. More broadly, our results highlight the utility of ultracold atomic superfluids as a platform for quantum simulation of nonequilibrium transport.
\section*{optical sample shaping}

We prepare a balanced mixture of the lowest and third lowest hyperfine state of fermionic ${^6}$Li in the unitary regime. A total of $2N=\SI{2.7 \pm 0.1 }{}\times 10^5$  atoms are cooled to $T/T_F = \SI{0.144 \pm0.007}{}$, where $N$ denotes the number of atom pairs, $T_F$ denotes the Fermi temperature and the errors denote the statistical uncertainties. The atoms are confined in a cigar-shaped, harmonic trap with trapping frequencies \mbox{$(\nu_x,\nu_y,\nu_z) = (\SI{222 \pm 7},\SI{28.1 \pm 0.1}{},\SI{195\pm5}{})~\SI{}{\hertz}$}.

As depicted in Fig.~\ref{fig: fig1}a, a repulsive $\textrm{TEM}_{0,1}$ beam, propagating in $x$-direction, is aligned with the trap center. It divides the atomic cloud into two reservoirs, separated by a quasi-two-dimensional transport region that is \SI{51 \pm 1}{\micro \meter} long and provides confinement of $\nu_{\mathrm{2D} }= \SI{10.46 \pm 0.09}{\kilo \hertz}$  along $z$. Within the central $\SI{27\pm0.5}{\micro \meter}$ of this plane, the atoms are further confined in the $x$-direction to a channel of variable width $\mathrm{w}$. The channel is defined by a repulsive beam propagating along $z$, shaped by a digital micromirror device (DMD). It enables reproducible tunability of $\mathrm{w}$ from \SI{0.9\pm 0.5}{\micro \metre} to \SI{19.8\pm0.5}{\micro \metre}, where the uncertainty is dominated by and calculated from the resolution limit of our projection system (NA=0.53, $\lambda = \SI{532}{\nano\meter}$). Note that given the dimensions of our setup, we are not in the Josephson regime (see \ref{sisec:NoJO}). A third, attractive Gaussian beam (not shown) illuminates the transport channel and its connections to the reservoirs, locally tuning the chemical potential. 
A more detailed description of the setup can be found in reference~\cite{fabritius_irreversible_2024}.
\section*{Phenomenology}

At time $t=0$, we prepare a relative atom-pair imbalance between the two reservoirs, \mbox{$\Delta N/N = (N_\mathrm{l} - N_\mathrm{r})/(N_\mathrm{l} + N_\mathrm{r})$},
where $N_{\mathrm{l,r}}$ denote the number of atom pairs in the left and right reservoirs. 
An initial imbalance of $\Delta N/N \approx 0.1$ relaxes via an approximately constant initial current, followed by damped oscillations of the imbalance. 
The discharge is distinctly non-exponential, indicating a nonlinear response. 
The damping, and in particular its dependence on the channel width $\mathrm{w}$, carries information about the underlying microscopic mechanisms.

\begin{figure}
    \includegraphics[width=0.5\textwidth,angle=0]{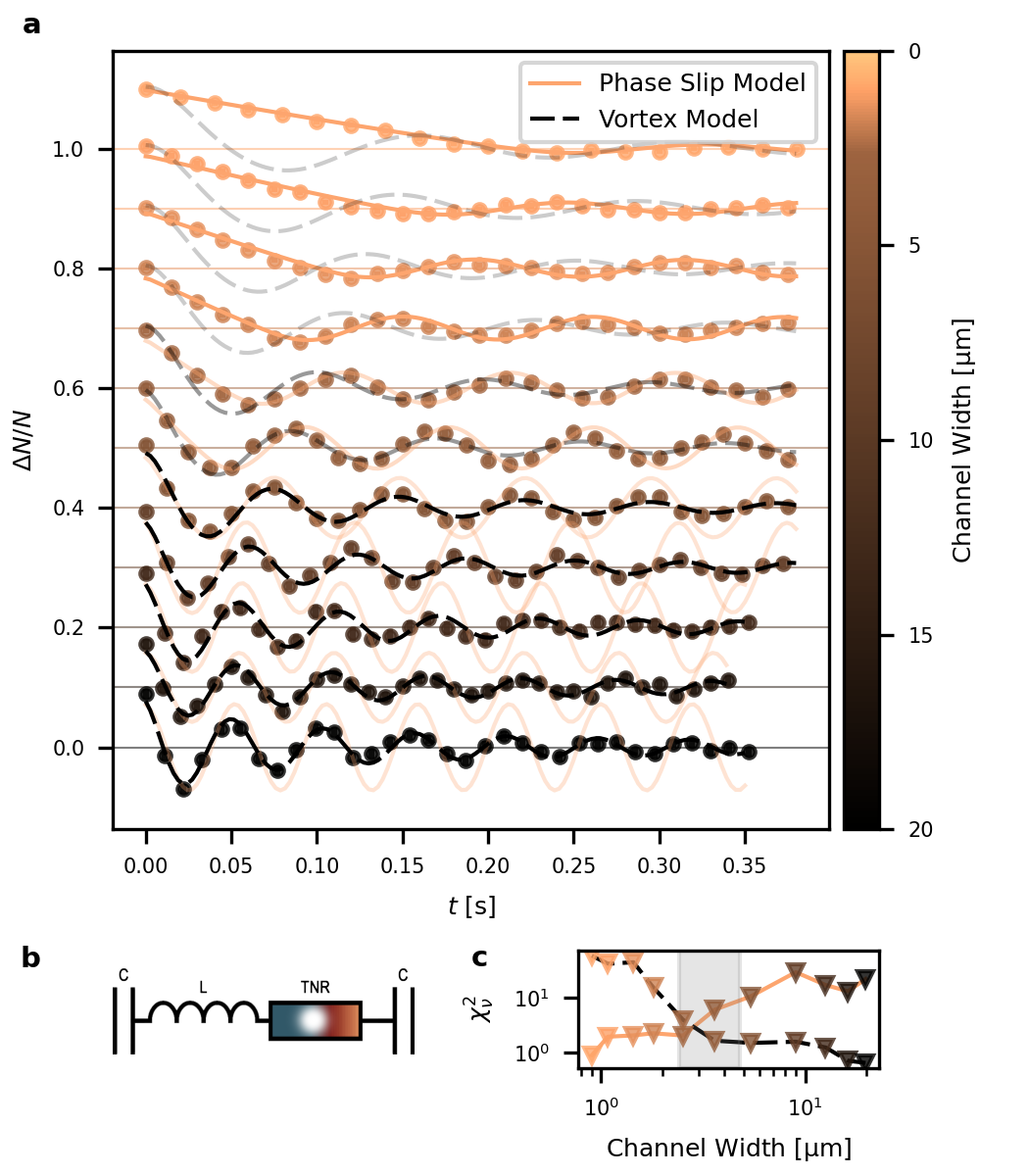}
    \caption{\textbf{Measured atom transport and fitted models}\newline
    \textbf{a}: Atom transport (imbalance $\Delta N/N$ as a function of time $t$) of a unitary Fermi gas through channels of increasing width shows distinctively changing dissipation characteristics. The data is compared to fitted phase slip (LAMH) and vortex models. The models both describe the Topological Nonlinear Resistor (TNR) in the circuit (\textbf{b}), capturing either phase slips or vortices. The LAMH model fits narrow channels, while the vortex model fits wide channels. The measurements for different channel widths are vertically offset for clarity. The statistical uncertainties of the measurements are smaller than the marker size.
    \textbf{c}: A quality of fit $\chi^2_\nu$ analysis substantiates the change of the dominant model apparent in panel a.}
    \label{fig: fig2}       
\end{figure}
The key data of our measurement series are shown in Fig.~\ref{fig: fig2}a. 
We record transport traces for $\mathrm{w}$ between $\SI{0.9\pm 0.5}{\micro\meter}$ and $\SI{19.8\pm 0.5}{\micro\meter}.$
As $\mathrm{w}$ increases, the magnitude of the initial current increases, which manifests in a steeper initial slope. 
Concurrently, both the oscillation frequency and amplitude grow. 
The frequency rises from just below \SI{10}{\hertz} to approximately \SI{18}{\hertz} (see also Fig.~\ref{fig: fig_A1}). 
For the narrowest channel ($\mathrm{w}=\SI{0.9\pm 0.5}{\micro\meter}$), the oscillation amplitude is comparable to the noise floor and barely resolved. 
In contrast, for the widest channel ($\mathrm{w}=\SI{19.8\pm 0.5}{\micro\meter}$), the amplitude reaches $\Delta N/N \approx 0.06$.

The damping exhibits a non-monotonic dependence on $\mathrm{w}$. 
At longer transport times (e.g.\ $t = \SI{0.3}{\second}$), the remaining oscillation amplitude reaches a maximum for intermediate channel widths (see also Fig.~\ref{fig: fig4}). 
These signals suggest distinct dissipative mechanisms in the narrow and wide channel limits, which we analyze quantitatively below.

\section*{Identifying Dissipation Mechanisms via Modeling}
The dynamics' general characteristics can be understood as arising from the interplay between the potential energy stored in the reservoirs, the kinetic energy inside the channel, and the dissipative processes we want to study. The compressibility of the gas allows for the definition of a capacitance $C_\mathrm{eff}$ (potential energy per atom pair displaced from equilibrium): \mbox{$\Delta E_\mathrm{pot}=1/(2C_{\mathrm{eff}}) \left(\Delta N/2\right)^2$}. In a first step, we measure the compressibility of our sample to determine $hC_\mathrm{eff}=\SI{15.8\pm 0.5}{\second}$, where $h$ is Planck's constant (See ~\ref{sisec:capacitance}). 
A charge neutral atom-pair current $I$ carries a kinetic energy that depends on the channel's width. Hence, we can assign a kinetic inductance $L(\mathrm{w})$ to the channel using $E_\mathrm{kin} = LI^2/2$. We model this inductance as $L=\ell_\mathrm{c}/\mathrm{w}+L_\mathrm{res}$, with fit parameters $\ell_\mathrm{c}$ and $L_\mathrm{res}$. The first term captures the kinetic energy inside the channel, the second term accounts for the residual kinetic energy outside the channel. While the mass $M$ inside the channel is proportional to its width $M\propto \mathrm{w}$, the velocity of the particles for a given current $v_I$ scales inversely $v_I\propto 1/\mathrm{w}$. Consequently, the kinetic energy inside the channel for a given current scales with $E_\mathrm{kin}=Mv^2/2\propto1/\mathrm{w}$. By analyzing the oscillation frequencies of all our measurements, we find $\ell_\mathrm{c}/h = \SI{3.00\pm 0.07 e-5}{\micro \meter\cdot\second}$ and $L_\mathrm{res}/h = \SI{3.06\pm 0.09 e-6}{\second}$ (See~\ref{sisec:inductance}).

Phase slips and vortices dissipate the current's kinetic energy into heat. So, they effectively act as a topological nonlinear resistor (TNR). These three components (potential energy in the reservoirs, kinetic energy, and dissipative processes in the channel) motivate the TNRLC model sketched in Fig.~\ref{fig: fig2}b. With $L$ and $C$ determined, the resistance $R$ remains the sole unknown component in our circuit. Owing to the nature of phase slips and vortices, $R$ is topological and non-linear, meaning it depends not only on the geometry but also on the current.

Uncovering how $R$ depends on geometry, in particular in relation to the length scale given by $\xi\approx\SI{1.2}{\micro\meter}$ (see~\ref{sisec:xi}), reveals the nature of the underlying dissipative processes.

\subsection*{Phase Slips in narrow Channels}
For narrow, quasi–1D channels ($\mathrm{w}\lesssim 2\xi$), we expect the system to be in the phase slip regime. 

The LAMH phase slip model predicts~\cite{doi:10.1142/S021797921005644X}
\begin{equation*}
     R(I,\mathrm{w})=\underbrace{R_0\exp\left(-\frac{\mathrm{w}}{\mathrm{w}_0}\right)}_{R_\mathrm{LAMH}(\mathrm{w})}\sinh\left(\frac{I}{I_0}\right)\frac{I_0}{I},
\end{equation*}

defined via $R=V/I$, where $R_0$ is a characteristic resistance, $\mathrm{w}_0$ a characteristic width, and $I_0=2 k_B T / h$ with $k_B$ the Boltzmann constant and $T$ the temperature. The exponential suppression of $R$ with $\mathrm{w}$ reflects the thermally activated phase slip rate scaling with $\Gamma\propto\exp(-\Delta F/(k_BT))$,\cite{PhysRevB.70.214506} where the free energy barrier for phase slips is $\Delta F\propto\mathrm{w}$.  The $\sinh(I/I_0)$-term can intuitively be understood as follows: At small $|I/I_0| \ll 1$, phase slips occur in both directions with roughly equal probability; as the bias increases slightly, phase slips that relax the current become more probable. For large $|I/I_0|\gg1$, $\sinh(I/I_0)\approx \exp(I/I_0)/2,$ reflecting the local destabilization of the superfluid and the resulting exponential increase in the phase slip probability.\newline Using our predetermined $C$ and $L(\mathrm{w})$, we determine $R_0$ and $\mathrm{w}_0$ by fitting them simultaneously to the measurements of our four narrowest channels (see Sec.~\ref{sisec:fitting}). We find $R_0/h = 10^{2.2 \pm0.2}$ and $\mathrm{w}_0 = \SI{50.2\pm0.6}{\nano \meter}$, where the errors denote the uncertainties of the fit. The resulting fits (shown in Fig.~\ref{fig: fig2}a) are of good quality ($\chi_\nu ^2< 2.3$) for narrow channels up to $\mathrm{w}=\SI{2.5}{\micro \meter}$, and degrade rapidly where the model is expected to break down.\newline In a second step, we validate the choice of the LAMH model by fitting $R_{\mathrm{LAMH}}(\mathrm{w})$ to each curve individually. Within the model’s validity regime, the individual fits reproduce the predicted exponential scaling over more than ten orders of magnitude, as shown in Fig.~\ref{fig: fig3}a. This is the first main finding of this paper. Compared to $R_{\textnormal{LAMH}}(\mathrm{w})$, the total resistance $R(I,\mathrm{w})$ apparent in the measurements (Fig.~\ref{fig: fig2}a) varies much less. As we initialize a fixed potential-bias in our system, the exponential geometric suppression ($R \propto \exp(-\mathrm{w}/\mathrm{w_0})$) approximately cancels the current-induced exponential enhancement ($R \propto \sinh(I/I_0) \approx \exp(I/I_0)/2$). This allows $R_{\mathrm{LAMH}}$ to be inferred over a wide range - more than 10 decades - while only exploring a relatively modest dynamic range of $R$. Importantly, the observed scaling of $R_{\mathrm{LAMH}}(\mathrm{w})$ is contained within the data itself and is not just an artifact of the choice of the model. This becomes evident when the model is applied outside its validity regime, where the scaling breaks down immediately (see Sec.~\ref{sisec:breakdown}).
\subsection*{Vortices in wide Channels}
\begin{figure}
    \includegraphics[width=0.5\textwidth,angle=0]{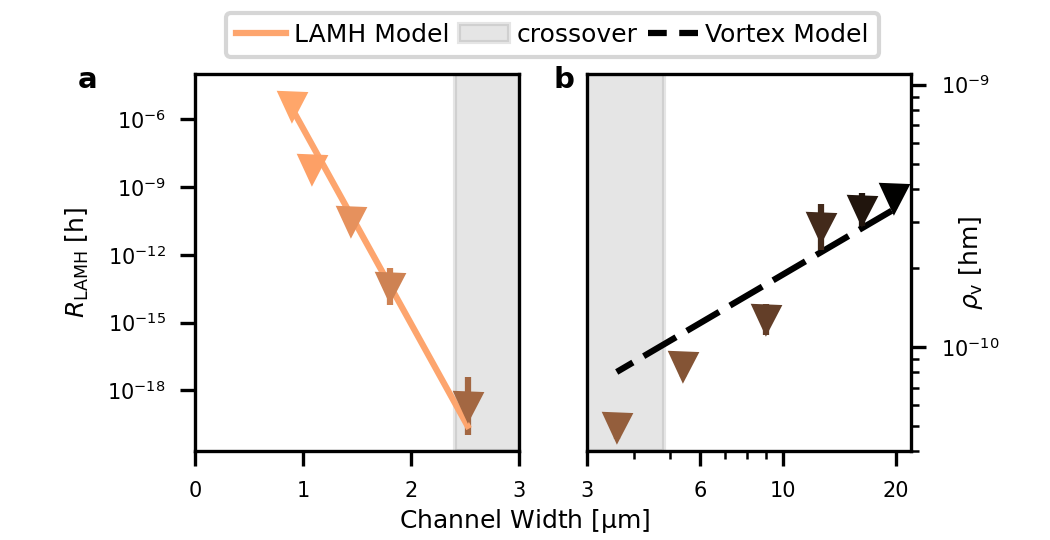}
    \caption{\textbf{Fitted resistances validate scalings laws}\newline \textbf{a}: The fitted phase slip resistance $R_{\textrm{LAMH}}(\mathrm{w})$, for narrow channels, shows the exponential suppression with width, as predicted by LAMH theory. \textbf{b}: The fitted vortex resistivity $\rho_{\mathrm{v}}(\mathrm{w})$, for wide channels, shows the predicted power-law. Inside the crossover region ($2\xi-4\xi$), i.e. outside the validity range of the models, both scalings break down (see Sec.~\ref{sisec:breakdown}). The errorbars represent the error of the fit.}
    \label{fig: fig3}       
\end{figure}

In quasi-2D channels $(\mathrm{w}\gtrsim 4\xi)$, phase slips become energetically unfavorable, as their activation energy increases exponentially with $\mathrm{w}$. Instead, dissipation is expected to be governed by vortices, the 2D analogue of phase slips.  Due to the vortices, we expect to see a power-law dependence in the pair-current density $J$: 
\begin{equation*}
    R(J,\mathrm{w})=\underbrace{\tilde{R}_0\xi\left(\frac{\mathrm{w}}{\xi}\right)^{\alpha-1}}_{\rho_{\mathrm{v}}(\mathrm{w})}\frac{1}{\mathrm{w}}\left(\frac{|J|}{J_0}\right)^{\alpha-1},
\end{equation*} with $J_0 = I_0/\xi$ (derivation in Sec.~\ref{sisec:vortexcalculation}).\cite{nakagawa_vortex_2024} By fitting $\alpha$ and $\tilde{R}_0$ simultaneously to the data of our four widest channels (see Sec.~\ref{sisec:fitting}), we find $\alpha = \SI{1.84 \pm 0.03}{}$, $\tilde{R}_0/h=10^{-4.57\pm0.06}$. As illustrated in Fig.~\ref{fig: fig2}a, this model captures the dynamics for channels in the 2D regime ($\chi^2_\nu\approx1$). In addition to the $\chi^2_\nu$ analysis presented in Fig.~\ref{fig: fig2}c, we validate our model by fixing $\alpha$ to its global fit value but fitting $\rho_{\mathrm{v}}$ to each individual curve. Within the validity range of the model, these individual fits agree with the predicted power-law $\rho_{\mathrm{v}}\propto \mathrm{w}^{\alpha -1}$ (see Fig.~\ref{fig: fig3}b). This scaling is the second main finding of the paper.
The scaling breaks down outside the model's validity regime (see Sec.~\ref{sisec:breakdown}), providing evidence that the observed scaling is a feature of the data, not a mere artifact of the enforced power law $R\propto J^{\alpha -1}$. When approaching the dimensional crossover $\mathrm{w}\lesssim 4\xi$, while still largely capturing the dynamics, the model begins to slightly but systematically underestimate the amplitude of the oscillations at long times. Presumably, because the approximation of sufficiently large $\mathrm{w}$ used in the derivation of $R(J)$ ceases to hold (see Sec.~\ref{sisec:vortexcalculation}). 
The eventual breakdown of the vortex model at $\mathrm{w}\approx4\xi$ marks the end of the 2D regime: 
the vortex model should be valid only when vortex–antivortex pairs are available as excitations. Since each vortex has a radius of $\xi$, such an excitation requires $\mathrm{w}\gtrsim4\xi$. This leaves a gap between the validity ranges of the phase slip and the vortex model, where neither model provides a microscopically satisfactory description. In this gap, there is a dimensional crossover, where both phase slips and vortices compete.
\newline

\section*{Exploring the dimensional crossover} 
\begin{figure}
\includegraphics[width=0.5\textwidth,angle=0]{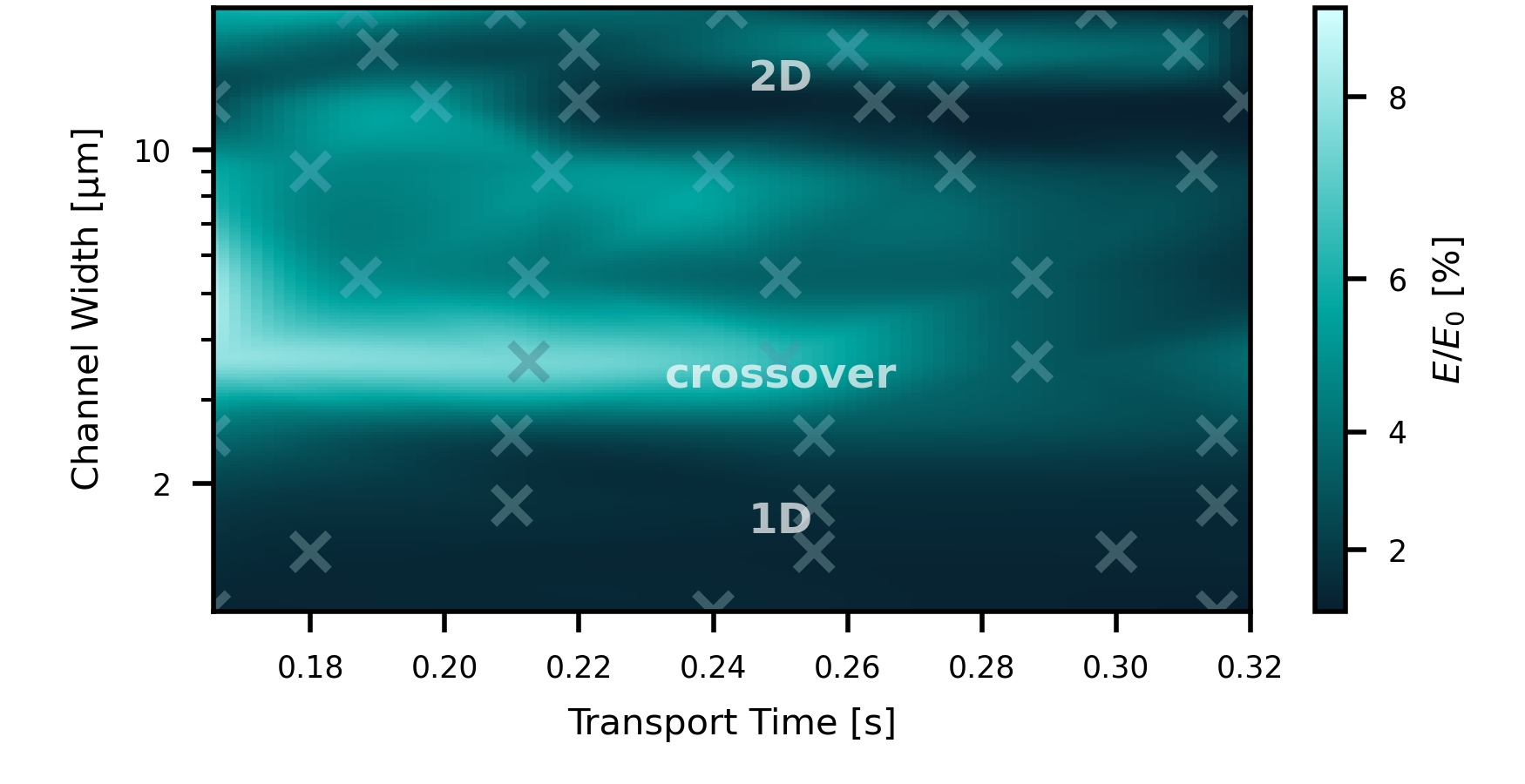}
    \caption{\textbf{Model free analysis reveals minimal dissipation in dimensional crossover} The remaining energy in the oscillator paradoxically varies non-monotonically as a function of the channels width $\mathrm{w}$. For a given time after the initial decay, the maximum appears in the crossover region, near $\mathrm{w}=\SI{4}{\micro \meter}\approx 3\xi$, where phase slips and vortices are suppressed but competing. The crosses mark the measurement points between which we interpolate.}
    \label{fig: fig4}       
\end{figure}

The underdamped oscillations allow us to characterize dissipation in our system using only a few basic assumptions, while staying agnostic to the dissipation mechanism. This is particularly useful to analyze the dimensional crossover, where neither the phase slip nor the vortex model provide a satisfactory microscopical description.
We split the total energy $E_\mathrm{tot}$ into a potential (capacitive) energy $E_C$ and a kinetic (inductive) energy $E_L$. We assume $E_\mathrm{tot}=E_C(\Delta N)+E_L(I,\mathrm{w})$ with $E_C\propto \Delta N^2$ and $E_L(I=0,\mathrm{w})=0$. For each channel width $\mathrm{w}$, we extract the times 
$t_{i,\mathrm{w}}$ at which $\Delta N(t=t_i,\mathrm{w})$ reaches an extremum and thus $I\approx0$. There, the system has minimal inductive ($E_L\approx0)$ and maximal capacitive energy. The fraction of the initial energy remaining in the system is then given by \begin{equation*}
\frac{E(t=t_{i,\mathrm{w}})}{E(t=0)}\approx\left[ \frac{\Delta N (t=t_{i,\mathrm{w}})}{\Delta N (t=0)}\right]^2.\end{equation*}
The remaining energy as a function of time and channel width is shown in Fig.~\ref{fig: fig4}. The plot is obtained by cubic interpolation between the measured points, performed linearly in time and logarithmically in width.

For a fixed width, the energy decays over time, reflecting dissipation. Strikingly, at any given time $t>\SI{0.15}{\second}$, $E_t(\mathrm{w})/E_0$ varies non-monotonically with the channel width, exhibiting a clear maximum near $\mathrm{w} \approx3 \xi$, corresponding to a minimal effective resistance. We call this deviation from the expected monotonicity the ''quantum resistance paradox'' and consider it the third and last main finding of this paper. Note that the activation energy for phase slips increases exponentially with channel width, while vortex–antivortex pairs require sufficiently wide channels to form. The observed non-monotonicity thus suggests a transition in the underlying dissipation mechanism and a crossover from 1D to 2D physics.

\section*{conclusion}
We have investigated atom transport between two bulk superfluids through channels across different dimensional regimes and observed distinct behaviors. In narrow, quasi-one-dimensional channels, our data are in agreement with the LAMH model. We infer the phase slips' geometric suppression over more than ten orders of magnitude, providing the first experimental confirmation of the predicted geometric scaling. For wide, quasi-two-dimensional channels, dissipation is captured by a vortex model.  The borders of the one- and two- dimensional regimes appear close to the theoretically expected length scale. A measurement of energy decay, agnostic to microscopic details, provides further evidence for a transition in the dissipation mechanism for channels of intermediate width, which paradoxically exhibit the least energy loss. These findings not only suggest routes for optimizing energy transport in mesoscopic devices but also provide a benchmark for theoretical efforts to describe the delicate dimensional crossover in superconductors.
\section*{Acknowledgements}
We thank A. Frank for his contributions to the electronics of the experiment. We are grateful for thorough discussions with J. Curtis, J. Marino, and S. Parameswaran.
We acknowledge the Swiss National Science Foundation (Grant Nos. 212168, UeM019-5.1, TMAG-2\_209376 and 232775) and ETH Zürich (Grant No. ETH-C-06 21-2) for financial support.
Y.Y. acknowledges support from the Council for Higher Education of Israel.

\section*{Author Contribution}
S.W., M.-Z.H., T.B., M.M., M.T. and Y.Y. upgraded the experiment and performed the measurements under the supervision of M.-Z.H.\newline S.W., A.G and F.M. identified the appropriate theoretical framework under the supervision of E.D.\newline S.W. analyzed the data (reviewed by M.-Z.H. T.B., M.M., M.T., and Y.Y.) and was in charge of writing the paper while all authors discussed the findings and contributed to the text and other aspects of the manuscript. M.-Z.H., E.D, and T.E conceptualized the project.
T.E. supervised the project.

\section*{Competing interests}
The authors declare no competing interests

\bibliography{references}

\clearpage

\section*{Supplemental material}
\setcounter{figure}{0}
\renewcommand{\thefigure}{\Roman{figure}}
\renewcommand{\tablename}{TAB.}

\section{Differential equations for RLC circuit and fitting procedure}
\label{sisec:fitting}
Our setup is analogous to an RLC circuit with the mapping presented in Tab.~\ref{table:analogy}:
\begin{table}[ht]
\label{table:analogy}
\centering
\begin{tabular}{c|c}
\hline
Superconducting RLC circuit & Our Setup \\
\hline
charge Q & atom-pair imbalance $\frac{\Delta N}{2}$\\
Cooper pairs & paired fermionic $^6\mathrm{Li}$ atoms\\
Capacitor & Reservoirs \\
Inductor & inertia of atoms \\
Resistance caused by & Resistance caused by \\
phase slips and vortices & phase slips and vortices \\
\hline
\end{tabular}
\caption{\textbf{RLC analogy} The table identifies the elements of a superconducting RLC circuit with their analogon in our setup.}
\end{table}
\noindent
The procedures for determining the capacitance $C$ and the inductance $L$ are described in Sections~\ref{sisec:capacitance} and \ref{sisec:inductance}. Here, we derive the governing equations and provide additional details on the fitting procedure used to determine $R$, as presented in the main text.

The total number of atoms in each reservoir is measured directly using absorption images of both hyperfine states ($\ket{1}$ and $\ket{3}$), taken in rapid succession. To determine the total number of pairs in each reservoir $N_{l,r}$, we use both measurements, as this reduces statistical uncertainty. Specifically, \mbox{$N_{l,r}=(N_{l,r;\ket{1}}+N_{l,r;\ket{3}})/2$}, where $N_{l,r;\ket{i}}$ denotes the number of atoms in the left/right reservoir in state $\ket{i}$. However, the second image (state $\ket{3}$) shows a cloud that has absorbed some energy from the first imaging pulse and is therefore slightly distorted. To avoid systematic effects from this distortion on $C$ and $T$, we use only the image of the lowest hyperfine state ($\ket{1}$) to extract the cloud's compressibility and temperature, assuming these are identical for both states. We verified that our preparation is equivalent for both states by inverting the imaging sequence. We describe our system in terms of pairs and report $C$, $L$, and $R$ for pairs and pair currents, respectively.

As shown in Fig.~\ref{fig: fig2}~b, our RLC circuit consists of two capacitors (reservoirs) in series. We replace these two capacitors $C$ by an effective capacitance $C_\mathrm{eff} = C/2$. With this replacement, we obtain a standard RLC series circuit with an initially charged capacitor but no externally applied voltage.

Applying Kirchhoff’s voltage law and charge conservation, we find:\begin{equation}
R I + L \frac{dI}{dt} + \frac{1}{C_\mathrm{eff}} Q = 0,
\end{equation}
\begin{equation}
\frac{dQ}{dt} = I.
\end{equation}

Using the mapping $Q \leftrightarrow \Delta N/2$ and $C_\mathrm{eff} = C/2$, we obtain:
\begin{equation}
\label{rlceq1}
\frac{dI}{dt} = - \frac{\Delta N}{2} \frac{1}{(C/2) L} - \frac{R}{L} I,
\end{equation}
\begin{equation}
\label{rlceq2}
\frac{d\Delta N}{dt} = 2 I.
\end{equation}

Phase slips and vortices induce nonlinear resistances, so $R = R(I)$ or $R=R(J)$, where $I$ and $J$ denote the pair current and pair-current density respectively. Depending on the model, we use:
\begin{equation}
    R(I) = R_0 \exp\left(-\frac{\mathrm{w}}{\mathrm{w}_0}\right) \sinh\left(\frac{I}{I_0}\right) \frac{I_0}{I},
    \end{equation}
    \begin{equation}
I_0 = 2 k_B T / h
\end{equation}
for the phase-slip model, and
\begin{equation}
R(J) = \tilde{R}_0\xi \left(\frac{\mathrm{w}}{\xi}\right)^{\alpha-1} \frac{1}{\mathrm{w}} \left(\frac{|J|}{J_0}\right)^{\alpha-1},
\end{equation}
\begin{equation}
J = I / \mathrm{w},\
\end{equation}
\begin{equation}
J_0 = I_0 / \mathrm{\xi}
\end{equation}
for the vortex model.

 The equations \ref{rlceq1} and \ref{rlceq2} are then integrated numerically.

Experimentally, transport is initiated by ramping down a blue-detuned beam that blocks the channel. A sudden quench of this beam slightly perturbs the reservoirs, so we instead ramp it down over \SI{10}{\milli\second}. This introduces an unknown initial condition $I(t=0)$. For each curve, we fit $I(t=0) \in [0, I_\mathrm{avg}]$, where $I_\mathrm{avg}$ is the average current during the initial decay (before oscillations). To avoid giving undue weight to the first measured point, we fit $\Delta N_{\mathrm{fit}}(t=0) \in [0.8, 1.2] \cdot \Delta N_\mathrm{measured}(t=0)$. These bounds are necessary to prevent the model from ignoring the initial decay, in particular for the wide channels, where we have only a few points describing the initial dynamics.

For numerical stability, $R_0$ and $\tilde{R}_0$ are fitted on a logarithmic scale, while $\alpha$ and $\mathrm{w}_0$ are fitted linearly.

Finally, we discuss the fitting weights. Each configuration $(t, \mathrm{w})$ has five measurements. Points with preparation errors (low atom number) are excluded, typically leaving four to five measurements per point and occasionally three. The experimental standard deviation is calculated for each configuration individually. We assume that for a given $\mathrm{w}$, the uncertainty is independent of $t$. To prevent the fit from being dominated by points with unrealistically small uncertainties, both the fitting and the $\chi_\nu^2$ calculations replace individual error bars with the median error bar of all the data sharing $\mathrm{w}$.

\section{Derivation of the Reservoir Capacitance}
\label{sisec:capacitance}

The capacitance of the reservoirs is fundamentally related to their compressibility.\cite{Krinner_2017} This connection stems from the analogy between the atomic transport system and an electrical RC circuit, where the compressibility serves as the neutral analog of electrical capacity. The compressibility of a single reservoir $\kappa$ is defined as

\begin{equation}
    \kappa =\frac{\partial N}{\partial \mu}\bigg|_T,
\end{equation}
where $N$ is the particle (in our case pairs) number, $\mu$ the chemical potential and $T$ the temperature of the reservoir.  
For the full transport system consisting of two reservoirs connected in series, the effective circuit capacitance is given by the series combination $C_{\text{eff}} = (C_L^{-1} + C_R^{-1})^{-1}$.
As mentioned in \ref{sisec:fitting}, we measure only the atom number in one hyperfine state, so we use $C\leftrightarrow \partial N_{\ket{1}}/\partial\mu =\kappa_{\ket{1}}$

\subsection{High-Saturation Imaging}
To determine the thermodynamic state, we use absorption imaging with intermediate intensities ($I/I_{\mathrm{sat}} \approx 1-2$) and a high-quantum-efficiency CCD camera.\cite{JeffThesis} The column density is reconstructed using the modified Beer-Lambert law to account for saturation effects:
\begin{equation}
    n_{\mathrm{col}}(y,z) = \frac{1}{\sigma_0} \left[ -\ln\left(\frac{I}{I_0}\right) + \frac{I_0 - I}{I_{\mathrm{sat}}} \right],
\end{equation}
where $\sigma_0 = 3\lambda^2/2\pi$ is the resonant scattering cross-section, $I$ is the transmitted intensity, $I_0$ is the incident intensity, and $I_{\mathrm{sat}}$ is the saturation intensity.

\subsection{Thermodynamic State Extraction via 2D EoS Fitting}
To determine the thermodynamic state of the reservoirs, we perform a two-dimensional fit to the reconstructed column density $n_{\mathrm{col}}(y,z)$. We assume the gas is in the unitary limit ($1/k_F a \to 0$) and in local thermodynamic equilibrium.

Under the Local Density Approximation (LDA), the local three-dimensional density $n(\mathbf{r})$ is determined entirely by the local dimensionless chemical potential $q(\mathbf{r}) = \mu(\mathbf{r})/(k_BT)$ via the universal function $f_n(q)$. In a harmonic trap, the local potential reads $q(\mathbf{r}) = q_0 - \sum (r_i/R_i)^2$, where $q_0$ is the central degeneracy and $R_i$ are the thermal radii.

The experimental signal is the column density integrated along the line-of-sight ($x$-axis). Integrating the universal density distribution yields the fitting function:
\begin{equation}
    n_{\mathrm{fit}}(y, z) = A \cdot N_0 \left( q - \frac{(y-y_0)^2}{R_y^2} - \frac{(z-z_0)^2}{R_z^2} \right) + \text{offset}.
\end{equation}
Here, $N_0(\tilde{q}) \equiv \int_0^\infty du \, u^{-1/2} f_n(\tilde{q}-u)$ is the calculated line-of-sight integral of the universal density function.

Although the ratio of the thermal widths $R_y/R_z$ is theoretically fixed by the trap aspect ratio $\omega_z/\omega_y$, we treat $R_y$ and $R_z$ as independent free parameters in the fit. This degree of freedom accounts for small drifts in the transverse confinement or slight anharmonicities in the optical potentials. This approach ensures that the temperature $T$, which is extracted primarily from the transport-axis width ($R_y$), remains robust against minor distortions in the transverse direction ($z$).

\paragraph{Extracting $T$ and $\mu$:}
The trap frequency $\omega_y$ along the transport direction is the most stable and harmonic parameter of the system. Therefore, the temperature is extracted primarily from the fitted width along the transport axis ($R_y$):
\begin{equation}
    T = \frac{m \omega_y^2 R_y^2}{2 k_B}.
\end{equation}
The chemical potential is then determined from the fitted degeneracy parameter:
\begin{equation}
    \mu = k_B T \cdot q.
\end{equation}
This procedure yields the properties of an ``extrapolated reservoir'' - the hypothetical full harmonic reservoir that would exist in the absence of our wall beam used to separate our harmonic trapping potential into two approximately half-harmonic reservoirs.

\paragraph{Atom Number Determination:}
While $\mu$ and $T$ are best determined from the fit to the harmonic wings, the total atom number $N$ is sensitive to the depletion caused by the wall. Therefore, rather than using the analytic integral of the fit (which would overestimate $N$, as it assumes the region depleted by the wall beam is filled with atoms), we compute $N$ by numerically integrating the measured column density over the region of interest:
\begin{equation}
    N_{L,R} = \int_{L,R} dy \int dz \, n_{\mathrm{col}}(y,z).
    \label{eq:atomNumber}
\end{equation}

\subsection{Calculation of the Capacitance}
With the thermodynamic coordinates $(q, T)$ of the reservoir established from the fit, we compute the capacitance.

It is important to distinguish between the \emph{absolute} atom number $N$ and the \emph{differential} capacitance $\kappa$.
For the absolute atom number (Eq.~\ref{eq:atomNumber}), we rely on numerical integration of the image because the presence of the wall beam approximately halves the relevant volume of the cloud.
However, for the capacitance ($\kappa = \partial N / \partial \mu$), performing a numerical derivative on the noisy integrated signal would yield an unstable result. Instead, we model the reservoir thermodynamically as a ``half-harmonic'' trap defined by the fitted parameters $(\mu, T)$. We assume that the intensive properties are determined by the bulk wings, while the extensive properties (like $N$ and $\kappa$) are exactly half that of a full harmonic trap due to the wall constraint.

To derive this analytic capacitance, we begin with the theoretical expression for the total atom number $N_{\mathrm{theo}}$ in a ``half-harmonic'' trap (integrating over only half the space). This introduces a factor of $1/2$ compared to the full harmonic result:
\begin{equation}
    N_{\mathrm{theo}} = \frac{1}{2} \left[ \frac{2}{\sqrt{\pi}} \left( \frac{k_B T}{\hbar \bar{\omega}} \right)^3 \int_0^\infty du \, \sqrt{u} \, f_n(q - u) \right],
\end{equation}
where $\bar{\omega}$ is the geometric mean trapping frequency.

The compressibility is then defined as the derivative of this theoretical atom number with respect to the chemical potential $\mu$ at constant $T$. We apply the chain rule using the relation $q = \mu /(k_B T)$, which implies $\frac{\partial}{\partial \mu} = \frac{1}{k_B T} \frac{\partial}{\partial q}$:
\begin{align}
    \kappa &= \frac{\partial N_{\mathrm{theo}}}{\partial \mu}\bigg|_T = \frac{1}{k_B T} \frac{\partial N_{\mathrm{theo}}}{\partial q} \\
    &= \frac{1}{k_B T} \left[ \frac{1}{\sqrt{\pi}} \left( \frac{k_B T}{\hbar \bar{\omega}} \right)^3 \right] \frac{\partial}{\partial q} \int_0^\infty du \, \sqrt{u} \, f_n(q - u).
\end{align}

The factor of $1/k_B T$ reduces the power of the temperature prefactor from 3 to 2. Furthermore, because the chemical potential enters only through the argument of the density function $f_n(q-u)$, the derivative $\partial/\partial q$ passes inside the integral and acts directly on $f_n$. This yields the final expression:
\begin{equation}
    \kappa = \frac{(k_B T)^2}{\sqrt{\pi}(\hbar \overline{\omega})^3 }\, N_0(q),
\end{equation}
where $N_0(q)$ is the dimensionless compressibility integral defined as:
\begin{equation}
    N_0(q) \equiv \int_0^\infty du \, \sqrt{u} \, f_n'(q - u).
\end{equation}
Here, $f_n'(x)$ denotes the derivative of the universal density function. The value of $N_0(q)$ is computed numerically using the known equation of state for the unitary gas at the experimentally determined degeneracy $q$.\cite{doi:10.1126/science.1214987}
\newline 
Note that for the analysis presented in the main text, the exact value of $C$ is of limited importance. A ratio $C_{\mathrm{measured}}/C_{\mathrm{true}}=r\neq1$ would only result in a global scaling of $L$ and $R$ by the same factor $r$. In particular, it would not affect the quality of the fits nor any conclusion on the dissipative mechanisms.

\section{Derivation of the Inductance}\label{sisec:inductance}
\begin{figure}
\includegraphics[width=0.5\textwidth,angle=0]{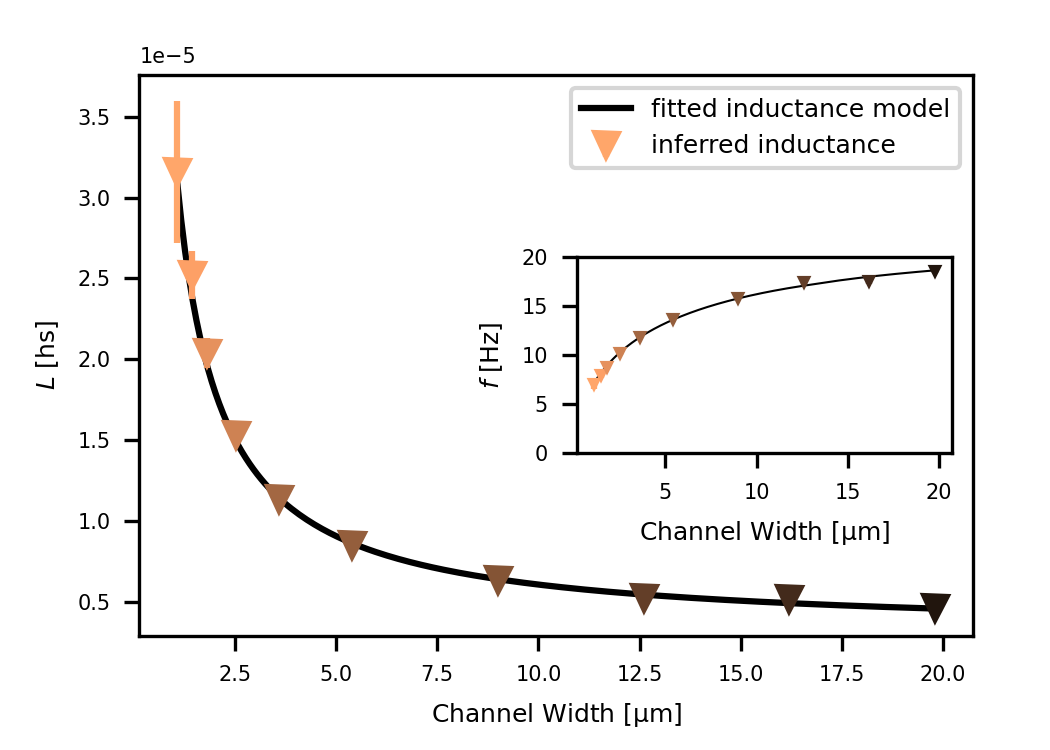}
    \caption{\textbf{Inductance as a function of channel width} The individually inferred inductances are well described by our fitted model. This model assumes a width dependent inductance of the channel in series with a width independent, residual inductance from the 2D region and the bulk superfluids. Inset: individually inferred oscillation frequencies are also well described by modeled
    frequencies relying on the same inductance model.}
    \label{fig: fig_A1}       
\end{figure}For all channel widths except the narrowest one, we observe underdamped, long-lived oscillations. Since these oscillations remain clearly underdamped, the influence of the nonlinear resistance on the oscillation frequency is expected to be small.
To extract the inductance of the channel for each width (except the narrowest one), we exclude the initial transient dynamics up to the first, global minimum occuring at a time 
$t=t_{1,\mathrm{w}}$. The remaining data are fitted with a damped sinusoidal function \mbox{$f_\mathrm{w}(t)=A_\mathrm{w}\exp(-(t-t_{1,\mathrm{w}})/\tau_\mathrm{w})\sin(\omega_ \mathrm{w}t+\phi_{\mathrm{w}})+\Delta N_{w,\mathrm{offset}},$} where $A_\mathrm{w},\tau_\mathrm{w},\omega_\mathrm{w},\phi_\mathrm{w},$ and $\Delta N_{\mathrm{w},\mathrm{offset}}$ are fitting parameters. Small offsets $|\Delta N_{\mathrm{w},\mathrm{offset}}|<0.018$ account for potential alignment errors that cause the oscillations to center around slightly nonzero values. Using the inferred capacitance $C$ of a single reservoir (see \ref{sisec:capacitance}), we determine the total inductance for a given channel width as $L_\mathrm{w}=2/((\omega_\mathrm{w}^2+\tau_\mathrm{w}^{-2})C).$ This kinetic inductance $L_\mathrm{w}$ quantifies the kinetic energy associated with a given pair current $I$. While the main contribution arises from the transport channel itself, there is also a residual contribution due to particle motion in the adjoining 2D and 3D regions, whose geometries are independent of the channel width. We therefore write $L(\mathrm{w}) = L_\mathrm{channel}(\mathrm{w})+L_{\mathrm{res}}$. Assuming a constant particle density inside the channel, the mass of particles in the transport channel ($M$) scales with its width, $M\propto \mathrm{w}$. For a given current, the corresponding particle velocity in the channel thus scales inversely, $v_\mathrm{channel}\propto\frac{1}{\mathrm{w}}$. Consequently, the kinetic energy associated with a fixed $I$ scales as $E_\mathrm{kin}=Mv_\mathrm{channel}^2/2\propto \mathrm{w} \left( 1/\mathrm{w}\right)^2 =1/\mathrm{w}$, implying that $L_\mathrm{channel}(\mathrm{w})\propto 1/\mathrm{w}$. We therefore model the total inductance as \begin{equation}L(\mathrm{w}) = L_\mathrm{channel}(\mathrm{w})+L_\mathrm{res} = \frac{\ell_\mathrm{c}}{\mathrm{w}}+L_\mathrm{res},\end{equation} with fitting parameters $\ell_\mathrm{c}$ and $L_\mathrm{res}$.  We find $\ell_\mathrm{c}/h = \SI{3.00\pm 0.07 e-5}{\micro \meter\cdot\second}$ and $L_\mathrm{res}/h = \SI{3.06\pm 0.09 e-6}{\second}$, where the uncertainties represent the standard errors of the fit.
The measured inductances $L_\mathrm{w}$ and the fitted model are shown in Fig~\ref{fig: fig_A1}. The excellent agreement of the model with the data validates our model assumptions.

\section{Estimate of \texorpdfstring{$\xi$}{xi}}
\label{sisec:xi}
In a previous, comparable experimental setting, we estimated the atoms traversing a region of length $d\approx$ \SI{3}{\micro \meter} in a time $t \approx$ \SI{330}{\micro \second}, resulting in an estimated atomic velocity of $v\approx$ \SI{9}{\frac{\milli \meter}{ \second}}.\cite{MohsenDarkState} Since our channel is practically defect free, we assume that the atomic velocity is close to the cricitcal velocity $v_c$. So we estimate $4\xi\approx4\hbar/(m v_c)\approx$ \SI{5}{\micro \meter}.\cite{RevModPhys.80.1215}

\section{Why we do not observe Josephson Oscillations}
\label{sisec:NoJO}
While several studies have reported on Josephson oscillations
(JO) with seemingly similar measurements in comparable
settings \cite{RoatiJosephsonBECBCS,RoatiJosephsonPhaseSlip,MoritzIdealJosephson}, our system operates far outside
the Josephson regime. The Josephson effect requires the
junction length to be much shorter than the superconducting
coherence length, which for our system is close to the healing length.\cite{Tinkham1996,doi:10.1142/S021797921005644X} In our case, the channel
length $\ell = \SI{27}{\micro \meter}$ is much larger than the estimated healing length $\xi\approx\SI{1.2}{\micro\meter}$ (See \ref{sisec:xi}).
Moreover, we observe neither a well-defined current–
phase relation nor an oscillation amplitude consistent
with the Josephson relation. We
therefore conclude that the oscillatory signal observed in
our experiment cannot be attributed to JO.

\section{Vortices in a finite width}
\label{sisec:vortexcalculation}
Here we outline the calculation of the current–voltage (I–V) characteristics of the XY-vortex model in a planar geometry that is periodic along the $x$ direction, with width $\mathrm{w}$ in $x$ and infinite extent in $y$.

We begin by considering the circulation of the gradient of the phase, which is given by
\begin{equation}
    2\pi n = \oint \boldsymbol{u}d\boldsymbol{s},
\end{equation}
where we parametrize the phase gradient $\boldsymbol{u}$ as a smooth and a topological part $\boldsymbol{u}=\boldsymbol{\nabla}\theta-\boldsymbol{\nabla}\cross(\boldsymbol{e}_z\psi)$ respectively. Here, $n$ is the total vorticity enclosed by the circulation. This can be parametrized as
\begin{equation}
    \boldsymbol{\nabla}^2\psi = 2\pi\sum_i n_i\delta^{(2)}(\boldsymbol{r}-\boldsymbol{r}_i),
\end{equation}
where $n_i\in \mathbb{Z}$ corresponds to the vorticity of the vortex at position $\boldsymbol{r}_i$. With this in mind, it is now clear that the energy due to vortex-vortex interaction can be obtain by the superposition principle once the Green's function is known. Therefore, we have to solve for
\begin{equation}
    \boldsymbol{\nabla}^2 G(\boldsymbol{r})=2\pi\delta^{(2)}(\boldsymbol{r})
\end{equation}
in the aforementioned geometry. This is easily achieved by employing the following Ansatz 
\begin{equation}
    G(\boldsymbol{r})=\frac{1}{\mathrm{w}}\sum_{m=-\infty}^\infty e^{i\frac{2\pi m}{\mathrm{w}}x}f_m(y),
\end{equation}
and after solving for each $m$, we arrive at
\begin{equation}
    G(\boldsymbol{r}) = -\frac{2\pi\abs{y}}{\mathrm{w}}-\sum_{m>0}\frac{e^{-\frac{2\pi m }{\mathrm{w}}\abs{y}}}{m}\cos\left(\frac{2\pi m}{\mathrm{w}}x\right).
\end{equation}
This equation can be simplified employing the following identity
\begin{equation}
    \sum_{m>0}\frac{z^m}{m}\cos(m\alpha)=-\frac{1}{2}\log(1-2z\cos\alpha+z^2)
\end{equation}
to obtain
\begin{equation}
    G(\boldsymbol{r})=-\frac{2\pi\abs{y}}{\mathrm{w}}+\frac{1}{2}\log\left[ \cosh\left(\frac{2\pi y}{\mathrm{w}}\right) - \cos\left( \frac{2\pi x}{\mathrm{w}} \right) \right].
\end{equation}
The energy due to the presence of vortices in this geometry is given by
\begin{multline}
    U(\boldsymbol{r})=-\frac{K}{2}\int d^2\boldsymbol{r}\,\psi\boldsymbol{\nabla}^2\psi = -2 \pi K\sum_{i,j}n_i n_j G(\boldsymbol{r}_i-\boldsymbol{r}_j),
\end{multline}
where $K$ is a constant related to the superfluid stiffness. We are interested in computing the production rate of vortex-antivortex pairs in the presence of an applied current along the $y$- direction, with current density $J$.

To compute the equilibrium separation between a single vortex-antivortex pair it is enough to set $y-y'=0$, such that the energy of the pair to be at a distance $x$ apart is
\begin{equation}
    U(x)=\pi K\log\left[ 1-\cos\left(\frac{2\pi x}{\mathrm{w}}\right) \right] - Fx,
\end{equation}
with $F~\sim J$ the Magnus force that arises due to the current going through the system. The distance $x^*$ at which the pair potential reaches its maximum corresponds to the location of the energy barrier that must be overcome to unbind a vortex–antivortex pair. It is obtained by extremizing the pair potential $U(x)$ with respect to $x$, i.e. by solving
\begin{equation}
    x^* = \frac{\mathrm{w}}{\pi}\cot^{-1}\left(\frac{F\mathrm{w}}{2\pi^2K}\right).
\end{equation}

This result holds as long as $x_{\rm eq}\gtrsim\xi$, the healing length. The energy barrier needed to overcome the confinement barrier is thus roughly given by $U(x_{\rm eq})$ and the unbinding rate $\Gamma\sim \exp(-U(x_{\rm eq})/T)\sim\left[ 1-\cos\left(2\cot^{-1}\left( \frac{F\mathrm{w}}{2\pi^2K} \right)\right) \right]^{-\pi K}\approx{(2/\left( \frac{F\mathrm{w}}{2\pi^2K} \right)^2)^{-\pi K}}$, where the last approximation is valid for sufficiently large $\mathrm{w}$. The voltage is then proportional to the current density and the number of free vortices. Since $\Gamma$ is the rate of vortex-antivortex pair unbinding, the numbers of free vortices in the system scales as $n_F\sim\Gamma^{1/2}$. Hence
\begin{equation}
    V\sim n_FJ\sim\Gamma^{1/2}J\sim\left(\frac{1}{\mathrm{w}J}\right)^{-\pi K}J = \mathrm{w}^{\pi K}J^{1+\pi K},
\end{equation}
where we have used that the Magnus force is proportional to the current density $F\sim J$.
Using Ohm's law, $V=RJ\mathrm{w}$, and substituting $\alpha = \pi K+1$ we find: \begin{equation}R\sim\mathrm{w}^{\pi K-1}J^{\pi K} = \mathrm{w}^{\alpha -1}\frac{1}{\mathrm{w}}J^{\alpha -1}.\end{equation}

\section{Breakdown of scaling laws}
\label{sisec:breakdown}

\begin{figure}
\centering
\includegraphics{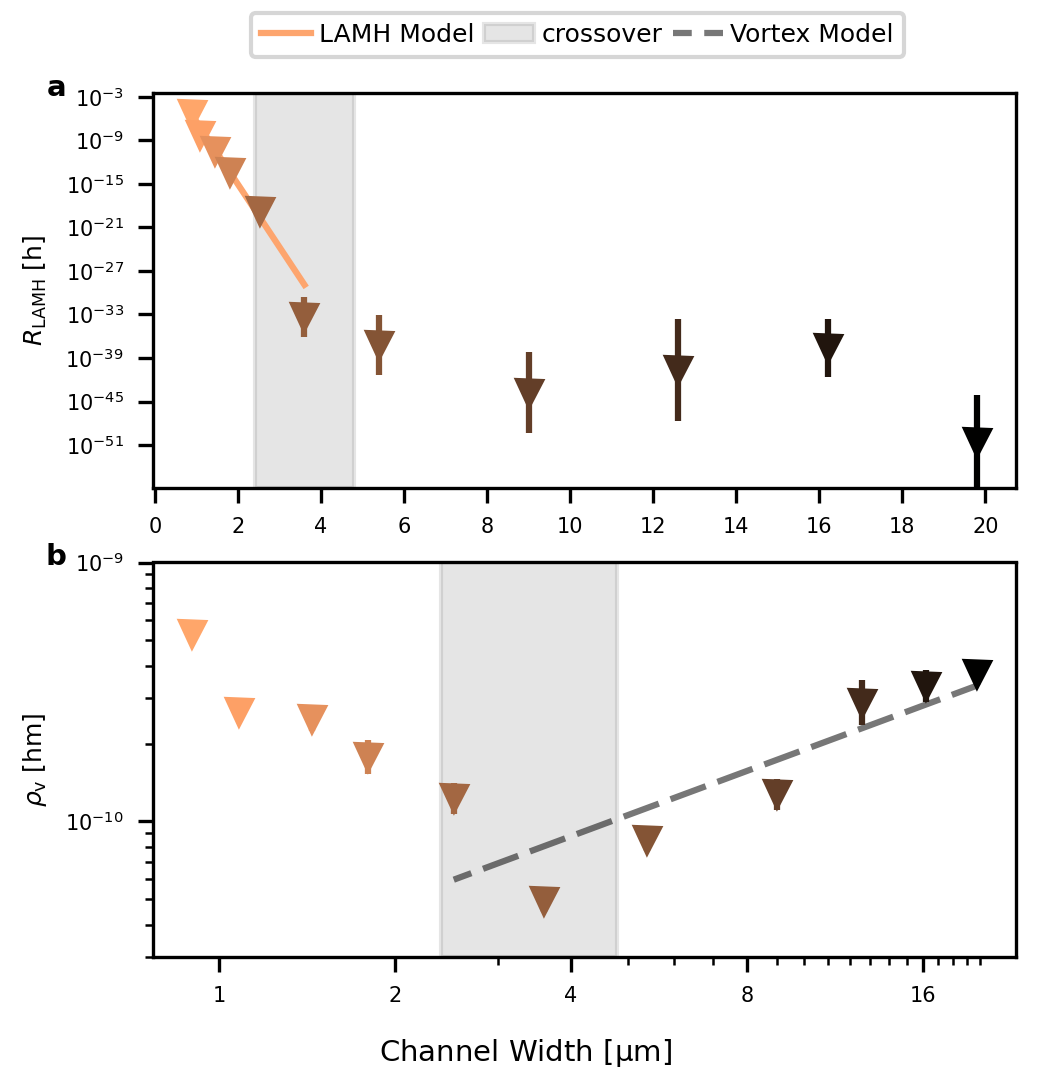}
\caption{\textbf{Breakdown of scaling laws at dimensional borders} Inferred resistances from the LAMH model (top) and the vortex model (bottom) are shown across the full width range. In both cases, the expected scalings break down outside the respective range of model validity. For the LAMH model, the inferred resistance does not encounter fitting boundaries or precision limits. Instead, the data exhibits a weak width dependence, consistent with the vortex model prediction. Portions of this data are also shown in Fig.~\ref{fig: fig3}.}
\label{fig: fig_A2}
\end{figure}

Figure~\ref{fig: fig_A2} provides additional data related to Fig.~\ref{fig: fig3}. The theoretical motivation and the $\chi_\nu^2$ analysis presented in Fig.~\ref{fig: fig2} already justify the choice of models. The data shown here complements that analysis by demonstrating that the scaling laws reported in Fig.~\ref{fig: fig3} are not numerical artifacts and are not imposed by the fitting procedure.

For both the LAMH and vortex models, the predicted scalings break down outside their respective ranges of validity. This behavior indicates that the observed scaling is not a numerical consequence of the fitting procedure and is not artificially enforced by the functional form of the models.

The saturation of the LAMH-inferred resistance at large widths does not seem to be  due to limited fitting precision, fitting boundaries, or lack of convergence. Instead, it reflects a transition from the strong exponential width dependence characteristic of the LAMH model to a much weaker dependence, as predicted by the vortex model.

For the vortex model, the trend reverses within the crossover regime, showing a clear deviation from the scaling expected from that model.

\end{document}